# Nanoscale fluorescence lifetime imaging with a single diamond NV center

Ryan Beams[1], Dallas Smith[1], Timothy W. Johnson[2], Sang-Hyun Oh[2], Lukas Novotny[1,3], A. Nick Vamivakas[1]

[1]Institute of Optics, University of Rochester, Rochester, New York 14627, United States

[2]Department of Electrical and Computer Engineering, University of Minnesota, Minneapolis, Minnesota 55455, United States

[3]ETH Zürich, Photonics Laboratory, 8093 Zürich, Switzerland

**Abstract:**

Solid-state quantum emitters, such as artificially engineered quantum dots or naturally occurring defects in solids, are being investigated for applications ranging from quantum information science and optoelectronics to biomedical imaging. Recently, these same systems have also been studied from the perspective of nanoscale metrology. In this letter we study the near-field optical properties of a diamond nanocrystal hosting a single nitrogen vacancy center. We find that the nitrogen vacancy center is a sensitive probe of the surrounding electromagnetic mode structure. We exploit this sensitivity to demonstrate nanoscale fluorescence lifetime imaging microscopy (FLIM) with a single nitrogen vacancy center by imaging the local density of states of an optical antenna.

Controlled interaction of light and matter at the nanoscale requires a detailed understanding of the local electromagnetic mode distribution. It has been recognized for some time that, in contrast to the homogeneity exhibited by the vacuum electromagnetic modes in free space, surfaces and nanostructured environments can sculpt the local electromagnetic mode density or local density of optical states (LDOS). A manifestation of this modification, as first pointed out by Purcell [1], is that the excited state lifetime of a quantum object is not an immutable property, but depends sensitively on the system's environment. Since the pioneering experiments of Drexhage [2] and Kleppner [3], numerous demonstrations have illustrated the impact LDOS engineering has on excited state decay dynamics [4-8]. Apart from a deeper understanding of how matter exchanges energy with the electromagnetic field, a detailed understanding of the LDOS is technologically relevant as photonic devices continue to shrink to the nanoscale [9].

One approach to LDOS mapping utilizes the excited state lifetime of an optically active material as a monitor of the local electromagnetic environment. During the last decade there has been a continued effort towards the realization of nanoscale fluorescence lifetime imaging microscopy (FLIM) with a single quantum system [10-16]. One system particularly well suited for this application is the nitrogen vacancy center (NV center) in diamond. NV centers have already found numerous applications ranging from quantum information science [17], single-photon generation [18,19] and quantum metrology [20-26] to fluorescence-based bioimaging [27]. Important for the present work is that the NV center exhibits stable photoluminescence at room temperature (unlike most quantum dots that suffer from blinking) and the center's optical properties are preserved when the diamond host takes the form of a nanocrystal. In the following we exploit the previous features to demonstrate, for the first time, the suitability of the NV center for nanoscale FLIM by imaging a nanoscale optical antenna [28].

The top panel of Fig. 1a presents an illustration of the FLIM concept. In the first modality a single quantum system (the "probe") is affixed to the vertex of a nanoscale tip and a sample to be interrogated (the "object") is scanned in close proximity to the emitter. The object perturbs the local electromagnetic environment of the probe and modifies its excited state dynamics that can be recorded as a function of the object's *x-y* coordinates to build an image. A second approach, followed in the present work, is to fix the probe in space, see bottom panel of Fig. 1a, and scan the object through the focus. Again, by monitoring the probe lifetime an image can be recorded. In our experiments a single NV center in a diamond nanocrystal is our probe of the nanoscale LDOS created by the object.

Figure 1b presents a schematic of our experimental setup. The diamond nanocrystal is prepared on a glass slide and mounted on a scanning *x-y* piezo stage in the focus of an inverted confocal microscope. The objective numerical aperture is 1.4. On top of the sample stage is a homebuilt AFM scan head that incorporates a silver (Ag) tip in the geometry of a pyramid (inset Fig. 1b, see Ref [29] for a discussion of the tip). The tip can be scanned independently of the sample in *x-y-z*. Either a continuous wave or pulsed 532-nm light source was used to excite the NV center. Prior to the objective the excitation beam is prepared in either radial or linear polarization. The resultant fluorescence is directed to either a spectrometer, a single avalanche photodiode (APD) for imaging (both far- and near-field fluorescence images) and lifetime measurements or to a pair of APDs for measurement of the intensity autocorrelation function. All measurements reported are done well below the saturation power of the NV center with the Ag tip contributing a background of ~ 4kHz.

A far-field scanning confocal fluorescence image under radially polarized illumination is presented in Fig. 2a. Radial polarization optimizes the near-field excitation of the NV center in the presence of the Ag tip. The spectrum in Fig. 2b corresponds to the location outlined with the white square in the confocal image. A measurement of the intensity autocorrelation function determines if the diamond nanocrystal hosts a single NV center. Figure 2c presents an exemplary autocorrelation function, without background subtraction, for the same NV center outlined in Fig. 2a. After determining a diamond nanocrystal hosts a single NV center the Ag tip was aligned to the confocal focus. The NV center was scanned through the Ag tip-focus region while fluorescence counts were recorded by the APD to construct a near-field image of the NV center. Figure 2d and 2f present near-field images of the same NV outlined in Fig. 2a (NV1) with 2 different Ag tips (tip1 and tip2). The corresponding topography maps, shown in Fig. 2e and 2g, allow us to determine the crystal heights to be ~ 20nm (the feature indicated by the arrow in Fig. 2e has a height of 10 nm). Shown as insets in both the near-field image and topography maps are linecuts along the direction indicated by the solid line. For the near-field image in Fig. 2d (2f) we observe a fluorescence enhancement of 10 (7). In all crystals studied we measured on average an order of magnitude fluorescence enhancement (9.7 ± 2.6). Not included in this average is the maximum enhancement we observed which was ~ 17 (see Supporting Information Fig. S1)

In comparing the two near-field images in Fig. 2 the most salient feature is the inhomogeneous fluorescence signal in Fig 2f. From the linecuts in Fig. 2d and 2e we observe the homogeneous fluorescence exhibits the same subdiffraction limited width (full width at half maximum (FWHM) of ~90 nm) as the crystal's topography. This should be contrasted with Fig. 2f where the fluorescence map exhibits localized hot spots (local increases in fluorescence counts) separated by ~ 90 nm with each feature revealing a subdiffraction limited width of 50 nm. This was not unique to this particular crystal-tip configuration (NV1/tip2) or excitation polarization. We observed local hot-spots on a number of other single-NV center containing nanodiamonds with different tips for both radial and linear excitation polarization (see Supporting Information Fig. S2 for another near-field image with hot-spots). A similar observation was reported for diamond nanocrystals containing multiple NV centers, but there it was not clear if the inhomogeneity was due to the presence of multiple emitters [15].

There are many possible effects that could result in local near-field hot spots. For example a double tip would manifest itself as two locations of increased fluorescence in the near-field image. We rule this out since the corresponding topography map (Fig. 2f) does not exhibit a similar inhomogeneity (the outline of the crystal topography is the closed white curve in Fig. 2f). A second cause is that for certain tip-crystal positions more light is scattered to the bucket APD detector. We do not think this is the cause since for the same tip-crystal position the excited state decay time of the NV center is reduced (see Supporting Information Fig. S3 for lifetime linecuts along the hot spots). A tentative explanation is the diamond nanocrystal operates as a dielectric antenna [30] and the tip as a metallic antenna. The two antennas have optimized arrangements that modify the LDOS seen by center optical transition. The appearance of the local hot spots then is a manifestation of specific tip-crystal configurations in which the combined effect of the tip and crystal antennas is to increase the LDOS and decrease the transition lifetime. These arrangements depend on the exact geometry of the tip and crystal and the same crystal will interact differently with tips exhibiting different geometries.

With an understanding of the near-field optical properties of the NV center we next study the impact an optical antenna has on the NV center radiative dynamics. Figure 3a and 3b present lifetime and intensity autocorrelation measurements on a single NV center with and without a proximal optical antenna (the top row in Fig. 3 correspond to NV1/tip1 and the bottom row are from NV2/tip3). In Fig. 3a (3b) the tip reduces the NV center lifetime by a factor of ~2 (~3) [Fig 3a. from 13 to 6 ns and Fig. 3b from 30 to 9.6 ns]. With this we conclude that the observed fluorescence enhancement is a combination of lifetime shortening and more efficient excitation of the NV center. To demonstrate nanoscale FLIM exploiting the NV-center optical-transition lifetime as a probe of an optical antenna's nanoscale LDOS the NV center is positioned in the confocal focus of the microscope, the Ag tip is raster scanned through the focus and for each tip $x$-$y$ location the lifetime is recorded. By working with the modality illustrated in Fig. 1a (bottom panel) we ensure the NV center excitation conditions do not change. Figures 3c and 3e and Figs. 3d and 3f present simultaneously acquired FLIM and near-field images for the two crystal-tip combinations (NV1/tip1 and NV2/tip3, white circles indicate crystal boundary from topography). The insets in Fig. 3c-3f are linecuts along the indicated lines. Again, comparing Fig. 3c and 3e we observe a localized hot spot in the FLIM (and near-field image) presented in Fig. 3e (Fig. 3f). Interestingly, the observed FLIM width in the case of Fig. 3e is ~ 30% narrower than the near-field fluorescence image width indicating the potential resolution increase afforded by FLIM as compared to near-field fluorescence imaging.

To illustrate the potential resolving power of the FLIM technique Fig. 4 presents $x$-, $y$- and $z$-linecuts of both the FLIM and near-field image of Figs. 3c and 3d (NV1/tip1). In the $x$-direction ($y$-direction) we find both approaches result in a full-width at half-maximum for the linecut of ~ 100 nm ( ~ 100 nm). We emphasize that the notion of resolution is complicated by the tip-crystal interaction, but one could imagine tailored nanocrystal geometries designed to exhibit responses that are better resolved than the length scale that characterizes the diamond crystal size. To complement the previous linecuts a final data set was recorded as the tip was retracted from the diamond nanocrystal. The measured decay length in this instance was ~ 10 nm. Combining the linecuts in the three perpendicular directions, the LDOS is probed in a volume corresponding to 100 x 100 x 10 nm$^3$.

In conclusion we have demonstrated that the NV center optical transition is a sensitive probe of the local electromagnetic mode structure. We observed an order of magnitude increase in the NV-center fluorescence and measured reduction in lifetimes of ~ 3. Although the diamond nanocrystal geometry influences the FLIM image, by fabricating crystals of specific geometry it will be possible to go beyond this proof-of-principle demonstration to make quantitatively accurate nanoscale LDOS maps. Additionally, the ability to resolve fluorescence and antibunching from a single NV center in the presence of an optical antenna suggests it should be possible to realize a scanning probe instrument to measure LDOS that affixes the diamond nanocrystal to the vertex of an optical antenna. The attractiveness of the NV center for LDOS imaging is that it could be used simultaneously to measure local electric and magnetic fields realizing a probe that could provide a wealth of nanoscale information in the vicinity of either target nanophotonic devices or biological material.

**Methods**

The diamond nanocrystals were obtained from Academia Sinica/Adamas. Single crystal samples on glass were prepared by first sonicating a dilution of diamond nanocrystals and then spin coating onto the glass substrate. The Ag tips were prepared as in [29]. To avoid oxidation, the Ag tips were kept in their silicon template until just before use. The laser excitations were a continuous wave 532-nm Coherent Compass and pulsed 100-ps 532-nm Picoquant. A single Perkin-Elmer avalanche photodiode was used for fluorescence images and lifetime images and a pair of APDs were used for intensity autocorrelation measurements. All correlations were done electronically with ORTEC nuclear instrument modules. The optical antenna was mounted onto a homemade AFM scan head and controlled by RHK electronics.

**Acknowledgements:** We thank P. Bharadwaj, M. Frimmer, Z. Lapin, and L, Rondin for valuable input and advice. A.N.V. acknowledges support from the Institute of Optics. L.N. acknowledges support from the Department of Energy (DE- FG02-05ER46207). T.W.J. and S.H.O. acknowledge support from the Office of Naval Research Young Investigator Award (N00014-11-1-0645) and the DARPA Young Faculty Award. D.S. acknowledges support from the NSF-IGERT (DGE-0966089).

Figure 1

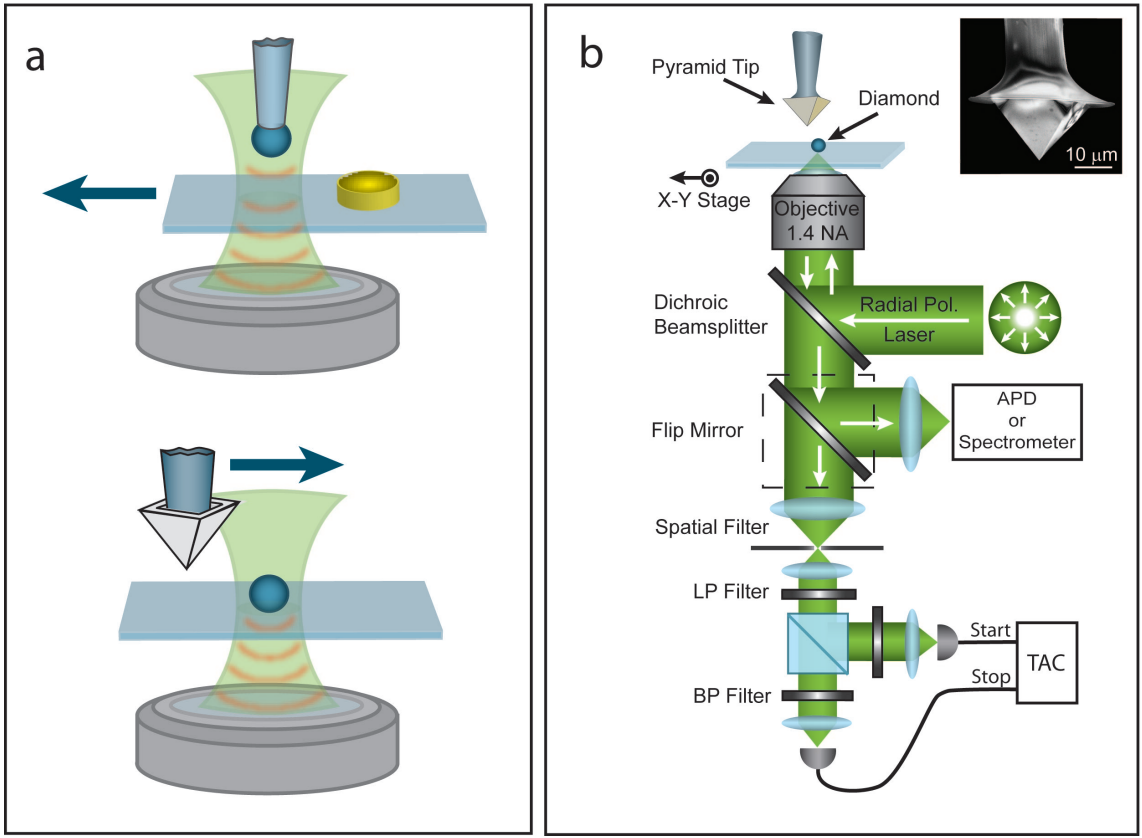

**Figure 1. Concept and experimental setup. a)** Top panel: Fluorescence lifetime imaging microscopy with a scanning quantum system. A target object is fixed in space and the probe quantum system is scanned in the vicinity of the target. The probe lifetime is recorded as a function of its position to build an image. Bottom panel: The probe is fixed in space and the target object is scanned in the vicinity of the probe to build a lifetime image. **b)** An inverted confocal microscope is combined with a homemade AFM scan head. The excitation beam polarization can be prepared in linear or radial polarization. The resultant fluorescence is spectrally resolved, collected by an APD for lifetime and fluorescence images, or sent to a pair of APDs for intensity autocorrelation measurements. LP = long pass filter. BP = bandpass filter. Inset: SEM image of a Ag pyramid (tip 1 in the text).

**Figure 2**

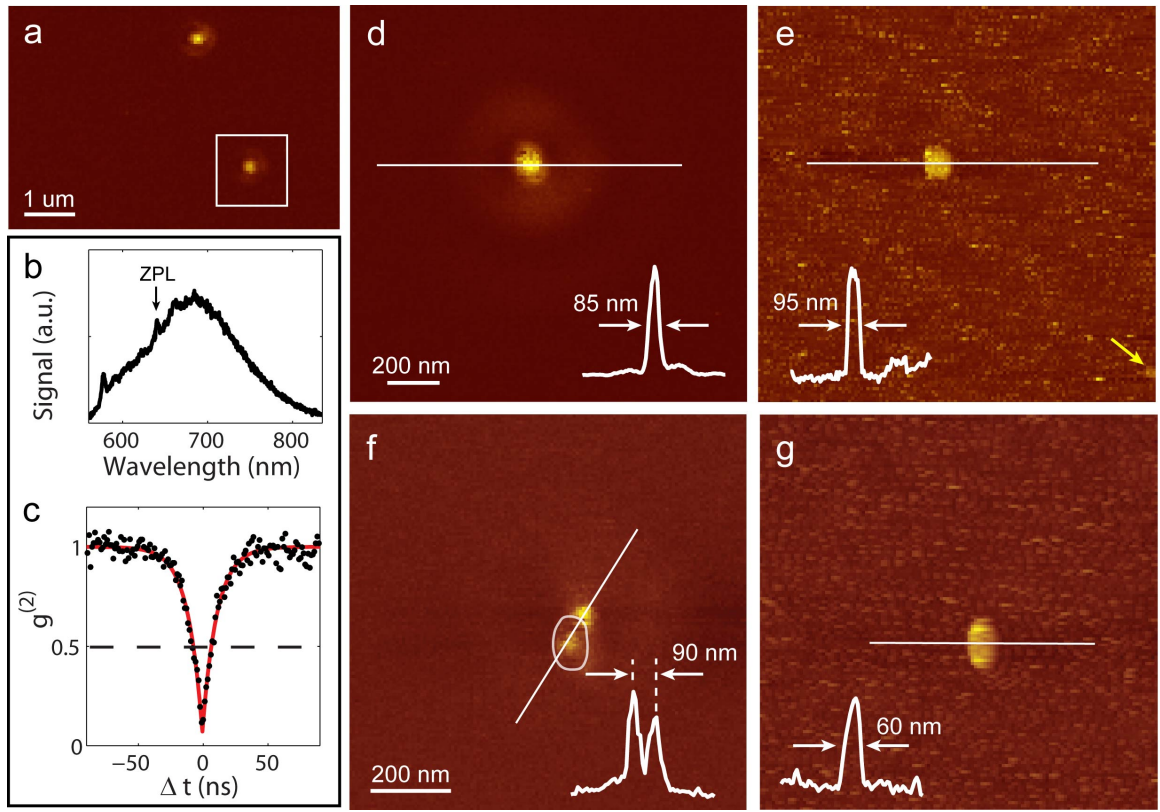

**Figure 2. Near-field optics of a single nitrogen vacancy. a)** Scanning confocal fluorescence image of NV centers under radial excitation. The white square denotes NV center 1. **b)** Spectrum of NV center 1. The zero phonon line (ZPL) is indicated. **c)** Intensity autocorrelation function of the fluorescence for NV center 1. There is no background correction. **d)** The near-field fluorescence image of NV center 1 with a Ag pyramid tip (tip 1). Inset: Linecut corresponding to the line indicated in the image. **e)** Topography map for NV center 1 with tip 1 taken simultaneously to the near-field image. Crystal height is ~ 20 nm. Arrow indicates small feature 10-nm high. Inset: Linecut corresponding to the line indicated in the image. **f)** The near-field fluorescence image of NV center 1 with a second Ag pyramid tip (tip 2). Inset: Linecut corresponding to the line indicated in the image. **g)** Topography map for NV center 1 with tip 2 taken simultaneously to the near-field image. Crystal height is ~ 20 nm. Inset: Linecut corresponding to the line indicated in the image.

**Figure 3**

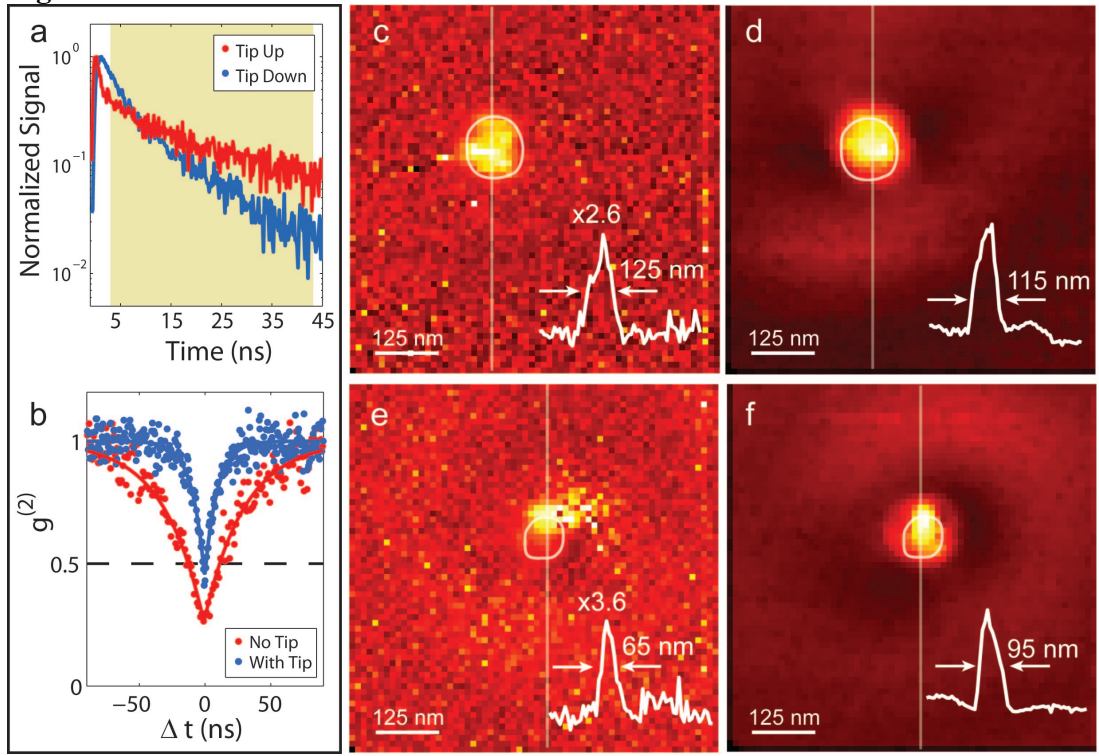

**Figure 3. Fluorescence lifetime imaging microscopy (FLIM) with a single NV center. a)** Lifetime measurements for NV center 1 with, and without, tip 1. The presence of the tip reduces the NV-center lifetime. The shaded region indicates the fitted region of the lifetime curve used in the FLIM image construction. **b)** Intensity autocorrelation function for NV center 2 and a third Ag tip (tip 3). The measurement is made with and without the tip. There is no background correction. **c)** Fluorescence lifetime image microscopy of tip 1 with NV center 1. The circle corresponds to the crystal topography. The inverse of lifetime is plotted (the decay rate). Inset: Linecut corresponding to the line indicated in the image. **d)** The near-field fluorescence image taken concurrent to the FLIM image in panel **c)**. Inset: Linecut corresponding to the line indicated in the image. **e)** Fluorescence lifetime image microscopy of tip 3 with NV center 2. The circle corresponds to the crystal topography. The inverse of lifetime is plotted (the decay rate). Inset: Linecut corresponding to the line indicated in the image. **f)** The near-field fluorescence image taken concurrent to the FLIM image in panel **e)**. Inset: Linecut corresponding to the line indicated in the image

**Figure 4**

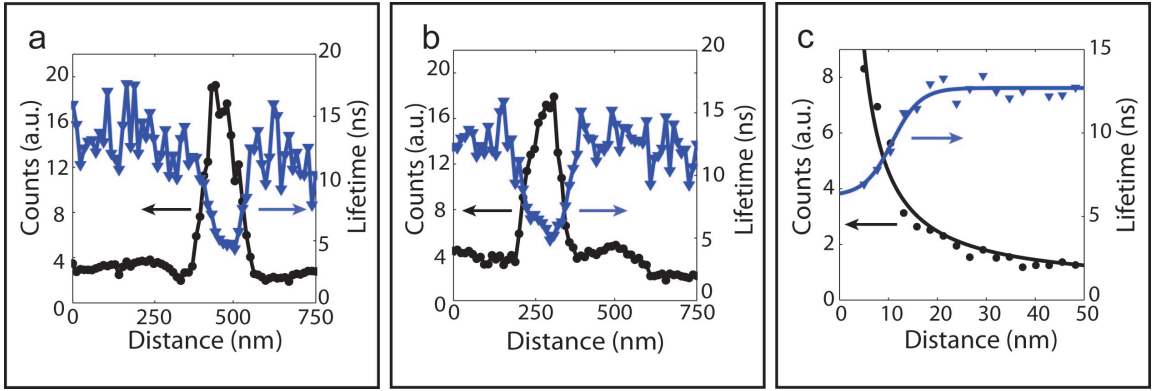

**Figure 4. FLIM resolving power. a)** Near-field fluorescence count rates and corresponding lifetimes for a *x*-linecut through the images in Fig. 3c and Fig. 3d. **b)** Near-field fluorescence count rates and corresponding lifetimes for a *y*-linecut through the images in Fig. 3c and Fig. 3d. **c)** Near-field fluorescence count rates and corresponding lifetimes for an approach curve taken with NV 1 and tip 1.

# Supporting Information

## 1. Maximum fluorescence enhancement

In the main text we report an average enhancement of 9.7±2.6 for all NV centers studied. This average does not include the maximum fluorescence enhancement we observed of 17. The near-field image of the crystal exhibiting the maximum enhancement is presented in Fig. S1a (the corresponding topography is in Fig. S1b). The inset of Fig. S1a is a linecut along the direction indicated. The enhancement is calculated by comparing the maximum confocal fluorescence counts with the maximum near-field counts. The 17 is a combination of excitation enhancement and increased radiative transition rate.

## 2. Near-field hot spots

An example of a near-field image of another single NV center that exhibits hot spots is presented in Fig. S2a. The curve in Fig. S2a is the outline of the crystal boundary as determined from the topography map in Fig. S2b. The near-field fluorescence is inhomogeneously distributed with respect to the diamond nanocrystal.

## 3. Lifetime linecuts of NV center 1 and Ag tip 1

The crystal discussed in Fig. 2f in the main text exhibits localized hot-spots in the near-field image. In Fig. S3a we present a second near-field image of NV center 1 with tip 2 from the manuscript. The lines in Fig. S3a indicate the linecuts presented in Fig. S3b-S3d. We observe at locations of increased near-field fluorescence counts the lifetime is correspondingly reduced.

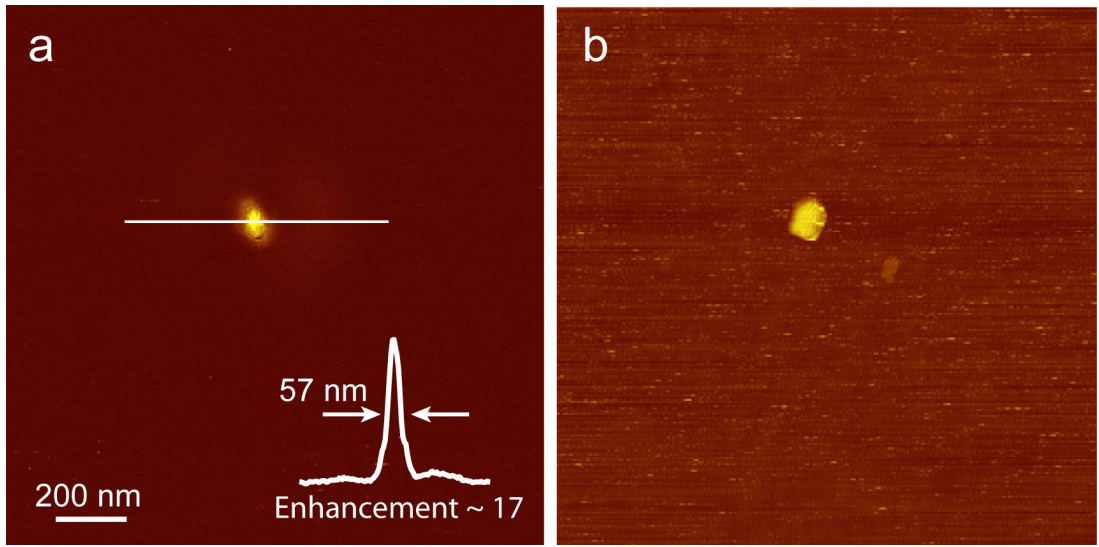

**Figure S1. Fluorescence enhancement. a)** Near-field image of a diamond nanocrystal with NV center. The fluorescence enhancement is determined to be 17 from comparison with the confocal fluorescence image of the same NV center. Inset: Linecut indicated by the solid line in the near-field image. **b)** Topography map taken simultaneously with the near-field image.

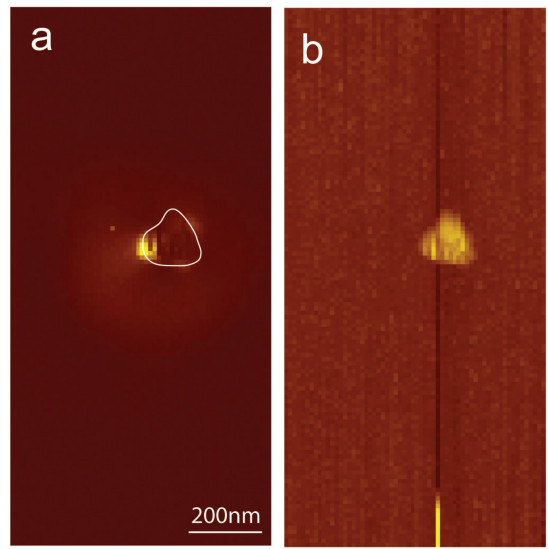

**Figure S2. Near-field fluorescence hot spots. a)** Near-field fluorescence image of an NV center. Inset: Linecut along line indicated in the near-field image. **b)** The topography map of the diamond nanocrystal.

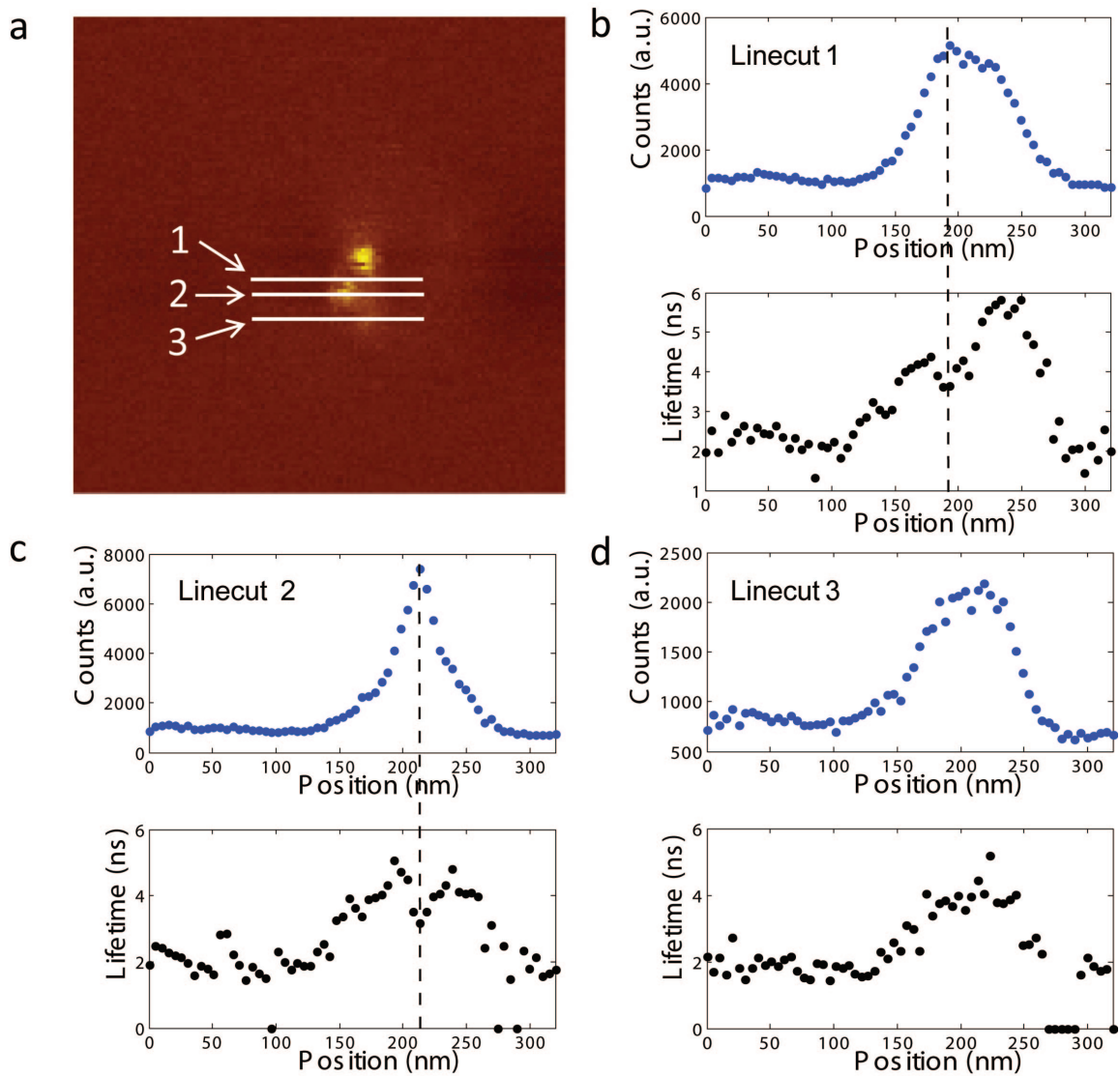

**Figure S3. Fluorescence lifetime linecuts. a)** Near-field image of the NV center in Fig. 2f and 2g. The solid lines indicate the linecuts presented in panels S3b-S3d. **b)** Near-field fluorescence and lifetime for the linecut indicated in Fig. S3a. Same for **c)** and **d).**

15